\documentclass[12pt,preprint]{aastex}
\usepackage{graphicx}

\def\xsrc{4U~0142+61~}
\def\xsrcnos{4U~0142+61}

\begin{document}
\vspace{0.8 in}

\title{Burst and Outburst Characteristics of Magnetar \xsrc}

\author{Ersin G\"o\u{g}\"u\c{s}\altaffilmark{1},
Lin Lin \altaffilmark{2},
Oliver J. Roberts\altaffilmark{3},
Manoneeta Chakraborty\altaffilmark{1},
Yuki Kaneko\altaffilmark{1},
Ramandeep Gill\altaffilmark{4},
Jonathan Granot\altaffilmark{4},
Alexander J. van der Horst\altaffilmark{5},
Anna L. Watts\altaffilmark{6},
Matthew Baring\altaffilmark{7},
Chryssa Kouveliotou\altaffilmark{5},
Daniela Huppenkothen\altaffilmark{8,9},
George Younes\altaffilmark{5}
}

\altaffiltext{1}{Sabanc\i~University, Orhanl\i-Tuzla, \.Istanbul 34956 Turkey}
\altaffiltext{2}{Department of Astronomy, Beijing Normal University, Beijing 100875 China}
\altaffiltext{3}{School of Physics, University College Dublin, Stillorgan Road, Belfield, Dublin 4, Ireland}
\altaffiltext{4}{Department of Natural Sciences, The Open University of Israel, 1 University Road, P.O. Box 808, Raánana 43537, Israel}
\altaffiltext{5}{Department of Physics, The George Washington University, Washington, DC 20052, USA}
\altaffiltext{6}{Anton Pannekoek Institute for Astronomy, University of Amsterdam, Postbus 94249, NL-1090 GE Amsterdam, the Netherlands}
\altaffiltext{7}{Department of Physics and Astronomy, Rice University, MS-108, P.O. Box 1892, Houston, TX 77251, USA}
\altaffiltext{8}{Center for Data Science, New York University, 726 Broadway, 7th Floor, New York, NY 10003, USA}
\altaffiltext{9}{Center for Cosmology and Particle Physics, Department of Physics, New York University, 4 Washington Place, New York, NY 10003, USA}

\begin{abstract}

We have compiled the most comprehensive burst sample from magnetar \xsrcnos, comprising 27 bursts from its three burst-active episodes in 2011, 2012 and the latest one in 2015 observed with \emph{Swift}/BAT and \emph{Fermi}/GBM. Bursts from \xsrc morphologically resemble typical short bursts from other magnetars. However, \xsrc bursts are less energetic compared to the bulk of magnetar bursts. We uncovered an extended tail emission following a burst on 2015 February 28, with a thermal nature, cooling over a time-scale of several minutes. During this tail emission, we also uncovered pulse peak phase aligned X-ray bursts , which could originate from the same underlying mechanism as that of the extended burst tail, or an associated and spatially coincident but different mechanism.

\end{abstract}


\section{Introduction}

Neutron stars with extremely strong magnetic fields (a.k.a., magnetars;~\cite{duncan1992}) are characterized by highly energetic, short (of ms duration) repetitive X-ray bursts during active episodes lasting days to months. Of the 29 magnetar candidates\footnote{http://www.physics.mcgill.ca/~pulsar/magnetar/main.html} currently known~\citep{olausen2014}, 24 sources have emitted bursts with peak luminosities close to/in excess of the non-magnetic Eddington limit. Burst repetition behavior varies significantly among magnetar candidates. Some magnetars emit tens, or even a few hundred bursts during an active episode~\citep{gogus2014}. Others emit only one or several bursts, usually coincident with the onset of rapid X-ray intensity increase (transient) episodes, which last for months or even years \citep{rea2011}.

According to the standard magnetar paradigm, bursts are the results of sudden fracturing of the neutron star crust under high magnetic pressure~\citep{thompson1995,thompson2001,lander2015}. Alternatively, magnetar bursts have also been suggested to be the result of magnetic re-connection~\citep{lyutikov2003}. For both scenarios, strong dipolar or multi-polar magnetic fields are expected. Recently identified magnetars with low inferred dipole magnetic fields, seem to be in conflict with the magnetar burst picture. For example, SGR~0418+5729 was found to have an inferred dipole field of 6$\times$10$^{12}$~G~\citep{rea2010, rea2012}. However, its surface magnetic field strength was determined to be 10$^{14}$~G~\citep{guver2011} via continuum X-ray spectral analysis, which is strong enough to trigger bursts. This finding was later confirmed by phase-resolved spectroscopy~\citep{tiengo2013}, indicating that much stronger field strengths are likely in multi-polar magnetic structures.

\xsrc is the brightest, persistent X-ray source among magnetars and a prominent emitter in hard X-rays~\citep{denhartog2008}, as well as in the optical and infrared~\citep{hulleman2004}. This is the only magnetar with a debris disk~\citep{wang2006}, however, it is still debated whether it is an active gaseous one~\citep{ertan2007} or a passive dust disk~\citep{wang2006}. \xsrc was once considered one of the most stable sources, emitting X-rays at a steady level~\citep{rea2007} and exhibiting a secular spin-down trend. Monitoring observations with the Rossi X-ray Timing Explorer (RXTE) revealed that \xsrc emitted energetic bursts in 2006 and 2007; the first activation is also associated with a sudden rotational frequency jump or timing glitch \citep{gavriil2011}. Bursts from \xsrc were highly unusual in the framework of typical magnetar bursts; two of them were extremely long (434, 1757~s) and their spectra showed peculiar emission features~\citep{gavriil2011}. Recently, \cite{chakraborty2016} re-analyzed the same data set and showed that these long events were bursts with extended tails, similar to those seen from SGR~1900+14~\citep{lenters2003}, SGR~1806$-$20~\citep{gogus2011}, and SGR~1550$-$5418~\citep{mus2015}. Time-resolved spectral analysis of these bursts using RXTE data also revealed variable but highly prominent X-ray absorption features around 6.5 and 11 keV, and an emission line at at $\sim13$ keV only during the very early episodes of their prolonged burst tails~\citep{chakraborty2016}.

\xsrc reactivated in July 2011 and January 2012, emitting bursts observed with \emph{Swift}~\citep{oates2011}. The source was burst-active again in February 2015, this time detected both by the \emph{Swift} Burst Alert Telescope (BAT)~\citep{barthelmy2015} and the Gamma-ray Burst Monitor (GBM)~\citep{roberts2015} on the \emph{Fermi} Gamma-ray Space Telescope. The bursts that triggered both BAT and GBM, were typically short events with most durations less than 0.1~s. In this study, we have performed deep searches in the archival BAT and GBM data to find additional events that were not luminous enough to trigger these instruments. We combined our results into the most extensive set of short magnetar bursts from \xsrcnos. Here we compare and quantify the spectral and temporal characteristics of these events, which appear to occur episodically every 0.5$-$3~years.

\section{2015 Reactivation}

\begin{figure}
\begin{center}
\includegraphics[scale=0.9]{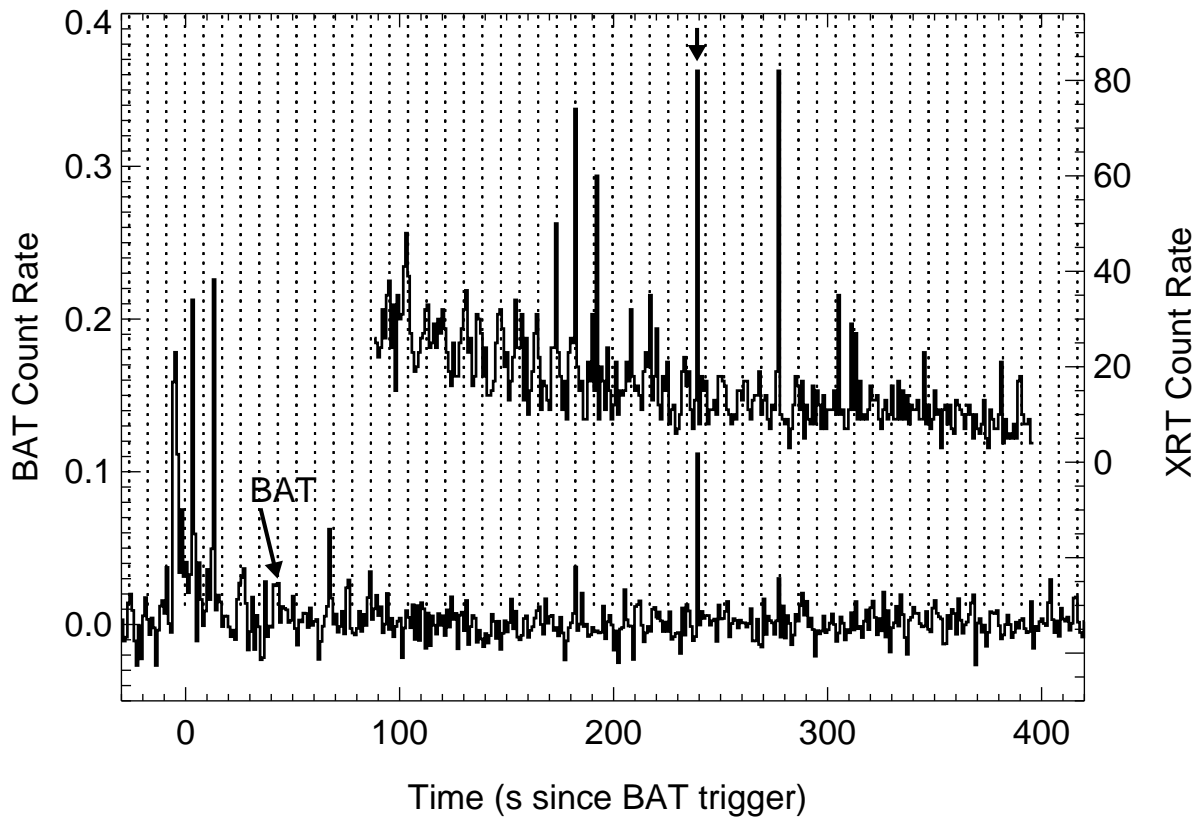}
\vspace{-0.0cm}
\caption{\emph{Swift}/BAT (left scale in the 15-150~keV band) observations of the leading burst, and XRT (right scale, 0.5-10~keV) observations of the following extended X-ray tail. The vertical dotted lines indicate the spin pulse peaks of the neutron star. The vertical arrow indicates the X-ray spike coincident with a short burst (see the text).}
\label{bat_xrt_lc}
\end{center}
\end{figure}

\emph{Swift}/BAT triggered on a burst from \xsrc on 2015  February 28, at 04:53:25 UT~(Barthelmy el al. 2015). The rapid slew of the spacecraft to the direction of the source resulted in follow up observations in Windowed Timing mode\footnote{This mode provides data with 1.7 millisecond time resolution without any significant pile-ip below 100 counts/s.} with the X-Ray Telescope (XRT) on-board \emph{Swift}, starting at $\sim$80~s after the BAT trigger. We show in Figure~\ref{bat_xrt_lc} the simultaneous \emph{Swift}/BAT and XRT lightcurves in 1\,s time resolution. The initial burst trigger was not captured with XRT, however, a decaying extended emission tail is observed, with superposed periodic X-ray modulations. To precisely determine the source spin period, we employed two additional XRT observations (2015 February 26; Observation ID: 00030738054, exposure of 4.1 ks, \& 2015 March 1; ID: 00030738055, exposure of 4 ks). We were able to establish a short term phase-connected spin ephemeris of the source covering the duration of the tail. Our timing solution yields a spin period P$_{\rm spin}$ = 8.68892(3) s. On Figure~\ref{bat_xrt_lc}  we indicate the peaks of the source spin phases as dotted vertical lines.

\begin{figure}
\begin{center}
\includegraphics[scale=0.9]{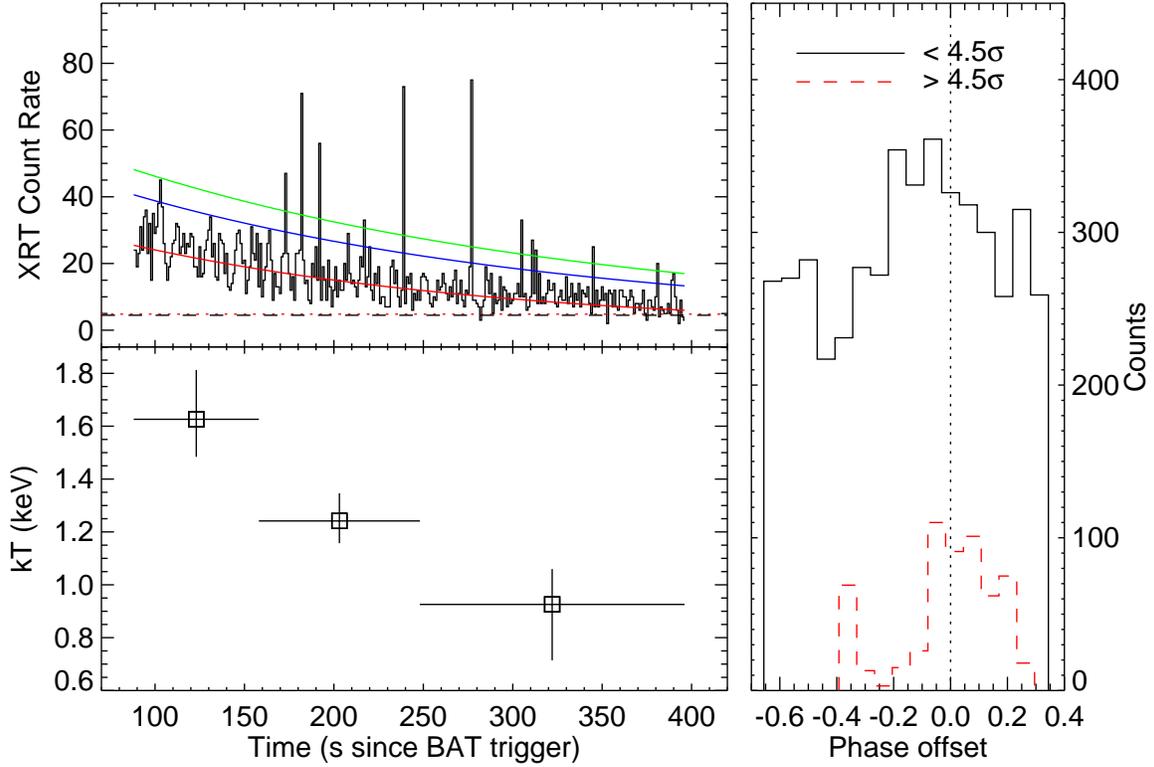}
\vspace{-0.0cm}
\caption{(Upper left panel) \emph{Swift}/XRT observations of the extended burst tail with 1 s time steps in the 0.5-10 keV band. The solid red curve is the exponential model fit, the blue and green curves are the 3.0, and 4.5 $\sigma$ levels above the decay trend, respectively. The horizontal red-dotted and black-dashed lines are the average X-ray count rates of \xsrc in the same energy band obtained from observations prior to and following the burst, respectively. (Lower left panel) The evolution of the blackbody temperature over the course of the extended burst tail. (Right panel) Phase distribution of XRT counts below and above the 4.5 $\sigma$ level in solid and dashed histograms, respectively. The former represent the pulse profile of the persistent emission without the spikes), and the latter is the phase distribution of the spikes/bursts. The vertical dotted line indicates the pulse peak.}
\label{xrt_kT}
\end{center}
\end{figure}

Further examination of the XRT lightcurve indicates the presence of sharp, short, intense bursts riding on the periodic X-ray modulations. The durations of these bursts do not exceed 100\,milliseconds\footnote{An accurate estimate of the burst durations was not performed, given the complexity of the intrinsic variability of the decaying tail trend.}, which is $\sim1\%$ of the pulse period. To determine the statistical significance of these bursts we estimate the average level of the decaying emission tail; we find that it follows an exponential trend with an initial rate of $38.5\pm1.6$ counts/s and an e-folding time of $212.4\pm8.4$\,s (red line in left upper panel of Figure~\ref{xrt_kT}). To estimate the tail duration, we compared its intensity level to the two XRT observations of \xsrc before and after the burst (mentioned above). We find an average X-ray count rate of $\sim4.5$ counts/s in both exposures, which is indicated with the overlapping horizontal dot-dashed lines in the upper left panel of Figure~\ref{xrt_kT}. We conclude that the X-ray tail emission had declined to the average pre- and post- burst level (within errors) by the end of the XRT pointing, thus constraining the total tail duration to $\sim300$\,s.

We now compare the position of the different structures (pulses and bursts) in the XRT lightcurve relative to the peaks of the pulse phase. We first estimate the 3.0, and 4.5 $\sigma$ levels above the average decay level (blue and green lines in left upper panel of Figure~\ref{xrt_kT}). We define all intensity levels larger than 4.5 $\sigma$ as bursts; we consider data below this level as part of the pulsed modulation. We then fold the XRT lightcurves both below and above the 4.5 $\sigma$ level. Figure~\ref{xrt_kT} (right panel) shows the two folded profiles: the upper closely reproduces the source pulse profile, while the lower exhibits the position of the bursts relative to the pulse peak phase. We note that the majority is within $-0.05$ to $+0.20$ of the pulse peak, with one exception at $-0.4$. The latter occurred at $T+238$\,~s, and is the only burst that has also been observed with the BAT (see Figure~\ref{bat_xrt_lc}). 

Although the XRT follow-up observation was short (about 300 s), enough X-ray data were acquired for performing a spectral analysis of the tail emission, due to the enhancement of the persistent X-ray emission induced by the bursts. To search for spectral evolution over the course of the tail, the spectrum was divided into three segments, each with a similar number of counts, defined as: Interval I (T+90~s to T+160~s), Interval II (T+160~s to T+240~s), and Interval III (T+240~s to T+390~s). The X-ray spectra extracted from all three segments, were simultaneously fit with an absorbed Black-Body (BB) plus Power Law (PL) model. As the interstellar hydrogen column density and power law index are not expected to vary over such a short duration, both parameters were linked so that they would converge to common values for all three spectra. During this process, we obtain a perfect fit ($\chi^2$/degrees of freedom = 215/214) yielding $N_{\rm H}$ = (1.2$\pm$0.2)$\times$10$^{22}$~cm$^{-2}$, and $\Gamma$ = 2.7$\pm$0.5. The temperature trend of the BB component is clearly shown to decline over the three spectra, with temperatures of 1.61$\pm$0.15~keV, 1.25$\pm$0.09~keV and 0.96$\pm$0.13~keV, measured for intervals I, II and III respectively (see the lower left panel of Figure~\ref{xrt_kT}). The corresponding X-ray flux of this thermal component over the 0.5$-$10 keV range for intervals I, II and III, was calculated to be (1.07$\pm$0.13)$\times$10$^{-9}$, (0.63$\pm$0.14)$\times$10$^{-9}$ and (0.20$\pm$0.10)$\times$10$^{-9}$~erg~cm$^{-2}$~s$^{-1}$, respectively. The radius of the BB emitting region remains constant (within errors) at $1.75\pm0.14$~km, (assuming the distance to the source to be 3.6~kpc; \cite{durant2006}). The normalization (i.e., the flux) of the power law component, which was allowed to float as a free parameter during the fitting process, remains constant within the determined errors. The spectra of the XRT observations two days before and a day after the enhancement are also suitably modeled with the absorbed BB plus PL, yielding $0.43\pm0.01$\,keV for the temperature of both intervals, and $3.64\pm0.07$ and $3.49\pm0.06$ for the PL indices before and after the reactivation episode, respectively. The pulsed fractions of the three tail intervals (I, II, III) were $0.18\pm0.03$, $0.30\pm0.03$ and $0.31\pm0.04$, respectively.

\section{Burst Observations and Sample}
\label{sect:obs}

\emph{Swift}/BAT (Barthelmy et al. 2005) and \emph{Fermi}/GBM~\citep{Meegan2009} are monitoring a large fraction of the unocculted sky in the hard X-ray/soft gamma-ray energy band, an optimal range for the acquisition of magnetar burst spectral data. Bright bursts from \xsrc triggered BAT and GBM in 2011, the BAT only in 2012, and both instruments again in 2015. However, not all magnetar bursts trigger the monitoring detectors due to intrinsically low intensity or instrumental constraints. Therefore, to obtain a complete list of bursts from \xsrc during its three active episodes, additional methods to extensively search the high time resolution continuous background data are required. One method uses Bayesian blocks, and the other searches for a minimal intensity excess over the local background. We briefly describe both techniques below, along with their results. 

The Bayesian blocks method represents the time-series data with step functions which correspond to maximum likelihood. It is not constrained by a priori amplitude or by the duration of the step functions~\citep{scargle98}. We used this method to find dim magnetar bursts in \textit{XMM-Newton} and \textit{Swift}/XRT observations~\citep{lin2013}. We applied our two-step search procedure to \textit{Swift}/BAT observations of \xsrc with two adjustments. Unlike photon counting instruments, the significance of BAT detections obeys a Gaussian distribution. Therefore, the first adjustment uses a likelihood function based on Gaussian statistics rather than Poisson statistics. Secondly, in order to focus on the signals from the source direction, we provided mask-weighted lightcurves to the search rather than the event lists. The lightcurve was extracted in the 15-150~keV energy band with 4~ms resolution and the box-car size was set to 4~s. A more detailed description of the search procedure can be found in~\citet{lin2013}. As a result of the aforementioned adjustments, our search found 13 additional bursts following the triggered event on 2011 July 29, and eight un-triggered bursts on the 2015 February 28 (which include three before the trigger). No additional events were found in the 2012 burst period. 

We also searched for magnetar bursts in the XRT data during the \textit{Swift}/BAT observations of the three burst-active episodes. Seven bursts were found in the XRT data with BAT counterparts, five in 2011 and two in 2015. Unfortunately, the XRT bursts were bright enough to suffer from pile-up, and were, therefore, not used in the joint-analysis. Table \ref{tab:batburst} shows the list of bursts detected by \textit{Swift}.

\begin{deluxetable}{cccccc}
\tabletypesize{\scriptsize}
\tablecaption{\textit{Swift}/BAT and \emph{Fermi}/GBM observations of bursts from \xsrcnos. \label{tab:batburst}}
\tablewidth{0pt}
\tablehead{
\colhead{Burst} & \colhead{Start time} & \colhead{T$_{Bayes}$} &
\colhead{T$_{90}$} &
\colhead{Detection} &
\colhead{Fluence\tablenotemark{**}} \\
\colhead{ID} & \colhead{in UTC} & \colhead{s} & \colhead{s} &
\colhead{ } & \colhead{}
}
\startdata
1\tablenotemark{T}	&	2011-07-29 11:19:15.398	&	0.008	&	&	BAT	& 0.28$\pm$0.03 \\
2	&	2011-07-29 11:19:38.918	&	0.008	&	&	BAT	& $0.7\pm0.1$\\
3	&	2011-07-29 11:20:17.026	&	0.024	&	&	BAT	& $0.8\pm0.2$	\\
4	&	2011-07-29 11:21:21.342	&	0.016	&	&	BAT	& $0.4\pm0.1$	\\
5	&	2011-07-29 11:21:33.082	&	0.008	&	&	BAT-XRT	& $0.4\pm0.1$	\\
6	&	2011-07-29 11:21:52.618	&	0.008	&	&	BAT	& $0.4\pm0.1$	\\
7	&	2011-07-29 11:21:57.830	&	0.012	&	&	BAT-XRT	& $0.7\pm0.1$	\\
8	&	2011-07-29 11:22:19.566	&	0.032	&	&	BAT	& $0.8\pm0.1$	\\
9	&	2011-07-29 11:22:24.894	&	0.004	&	&	BAT-XRT	& $0.2\pm0.1$	\\
10	&	2011-07-29 11:23:37.218	&	0.008	&	&	BAT	& $0.9\pm0.1$	\\
11	&	2011-07-29 11:25:05.058	&	0.008	&	&	BAT	& $0.3\pm0.1$	\\
12	&	2011-07-29 11:26:20.422	&	0.016	&	&	BAT-XRT\tablenotemark{p}	& $0.16\pm0.02$	\\
13	&	2011-07-29 11:28:31.274	&	0.008	&	&	BAT	& $0.5\pm0.1$	\\
14	&	2011-07-29 11:28:31.670	&	0.680	&	&	BAT-XRT	& $1.8\pm0.4$	\\
15	&	2011-07-29 17:40:37.124	&	--	&	0.020(6)$^{*}$ &	GBM 	& 12$\pm$1	\\
16\tablenotemark{T}	&	2012-01-12 13:09:38.665	&	0.028	&	&	BAT	& $0.5\pm0.1$	\\
17	&	2015-02-28 04:53:15.911	&	0.372	& Too weak	&	BAT-GBM	& $4.7\pm0.5$	\\
18	&	2015-02-28 04:53:18.383	&	0.036	&  Too weak	&	BAT-GBM	& $1.4\pm0.3$	\\
19	&	2015-02-28 04:53:20.323	&	0.044	& Too weak	&	BAT-GBM	& $1.8\pm0.3$	\\
20\tablenotemark{T}	&	2015-02-28 04:53:25.023	&	0.052	& 0.056(9)$^{*}$	&	BAT-GBM	& BAT: $6.9\pm0.5$ , GBM: 12$\pm$1 	\\
21	&	2015-02-28 04:53:35.195	&	0.036	& 0.030(8)$^{*}$	&	BAT-GBM	& BAT: $9.7\pm0.7$1 , GBM: 6$\pm$1 \\
22	&	2015-02-28 04:54:29.431	&	0.060	&	&	BAT	& $1.6\pm0.2$	\\
23	&	2015-02-28 04:54:37.643	&	0.172	&	&	BAT	& $2.7\pm0.3$ 	\\
24  &	2015-02-28 04:57:21.307	&	0.068	& 0.048(17)$^{*}$	&	BAT-XRT\tablenotemark{p}-GBM	& BAT: $6.2\pm0.4$ , GBM: 9$\pm$1 \\
25	&	2015-02-28 04:57:58.747	&	0.064	&	&	BAT-XRT\tablenotemark{p}	& $1.4\pm0.2$	\\
26	&	2015-02-28 05:06:55.645	&	--	& 0.070(22)$^{*}$	&	GBM	& 29$\pm$2	\\
27	&	2015-02-28 05:08:34.157	&	--	& 0.128(36)$^{*}$	&	GBM	& 3$\pm$1	\\

\enddata
\tablecomments{\\
$^T$ BAT triggered burst \\
$^p$ XRT observation is piled-up \\
$^{*}$ 8-200~keV, BB Spectral Model \\
$^{**}$ BB model fluence in units of 10$^{-8}$~erg cm$^{-2}$ in 15-150~keV for BAT bursts and 8$-$200 keV for GBM bursts. The fluence of burst 25 is from fitting a BB+BB model.}
\end{deluxetable}

\section{Temporal Properties of \xsrc Bursts}

The morphology of the bursts was determined by the Bayesian Block method~\citep{lin2013}. The burst duration is defined as the time from the start to the end of the burst blocks. The most important advantage of the Bayesian block duration is that the change point between the background and the burst is determined using an algorithm, and thus does not suffer from any of the bias that may occur as a result of using other techniques (e.g., selection of background interval). The Bayesian block durations of the BAT bursts are listed in Table~\ref{tab:batburst}. 

The T$_{90}$ duration for all of the bursts from \xsrc (defined as the duration during which the background-subtracted cumulative count rate increases from 5~\% to 95~\% of the total counts;~\citet{kouveliotou1993}), were determined in a manner similar to the method described by~\citet{lin2011}. The duration of the bursts were calculated using continuous time tagged event (CTTE) data of GBM, and the RMFIT\footnotemark \footnotetext{http://fermi.gsfc.nasa.gov/ssc/data/analysis/rmfit/} (v4.4.2) software, similar to what was done for GBM GRBs~\citep{paciesas2012} and other SGR events~\citep{vonkienlin2012}. The CTTE data type allows finer time bins to be generated which is necessary for the temporal analysis of short bursts from magnetars. The individual burst data were re-binned to 2, 4 or 8~ms depending on the intensity of the event. We present the T$_{90}$ duration for all triggered and un-triggered GBM bursts from outbursts in 2011 and 2015 in Table~\ref{tab:batburst}. Note that in determining the duration, we used a BB model over an energy range of 8$-$200~keV. We only used data from the \emph{Fermi}/GBM NaI(Tl) detectors with source to detector zenith angle $\theta \leq$40$^{\circ}$, for our temporal and spectral analyses.   

\section{Spectral Properties of the \xsrc Bursts}

The data for the bursts presented in Table~\ref{tab:batburst}, were fit with three continuum models which are known to best approximate the spectra of magnetar bursts: a BB model, a combined BB+BB model, and a PL function with an exponential high-energy cutoff (also known as the Comptonized model or in short COMPT). The RMFIT software package was used to analyze the spectral properties of the bursts detected with \emph{Fermi}/GBM, and XSPEC v12.9 for those bursts detected with \emph{Swift}/BAT. The aforementioned continuum models were first fit individually to the GBM burst data over an energy range of 8-200~keV, and 15-150~keV for BAT. Only the spectral properties of bright bursts could be investigated, and thus the intrinsically fainter bursts could not be used for subsequent spectral analysis. Table~\ref{tab:spec} lists the spectral model parameters resulting from the fitting of the three continuum models to bright bursts during the burst-active episodes in 2011, 2012 and 2015.

\begin{deluxetable}{cc|cc|ccc|ccc}
\tabletypesize{\scriptsize}
\tablecaption{Spectral burst properties of 4U~0142+61~\label{tab:spec}}
\tablehead{
Burst & Instument & \multicolumn{2}{c|}{BB} & \multicolumn{3}{c|}{BB+BB} & \multicolumn{3}{c}{COMPT} \\ 
ID  &      & kT & $\chi^2$/DOF & kT$_1$ & kT$_2$ & $\chi^2$/DOF & $\alpha$ & E$_{\rm p}$ & $\chi^2$/DOF \\
    &      & (keV) &   & (keV) & (keV) &   &  & (keV) &  
}
\startdata
 1 & BAT & $10.4\pm1.0$ & 14.5/21 &  &  &   & $0.6\pm1.2$ & $39.8\pm4.4$ & 13.7/20 \\ 
 2 & BAT & $5.7\pm1.5$ & 9.7/10 &  &  &   &  &  &  \\ 
 3 & BAT & $16.7\pm3.9 $ & 13.6/9 &  &  &   &  &  & \\ 
 4 & BAT & $9.2\pm2.3$ & 12.1/8 &  &  &   &  &  &  \\ 
 5 & BAT & $15.7\pm3.0$ & 7.1/10 &  &  &   &  &  &  \\ 
 6 & BAT & $13.2\pm2.8$ & 4.1/9 &  &  &   &  & &  \\ 
 7 & BAT & $11.9\pm1.7$ & 12.2/14 &  &  &   &  &  &  \\ 
 8 & BAT & $12.4\pm2.6$ & 6.8/7 &  &  &   &  &  &  \\ 
 9 & BAT & $4.4\pm0.5$ & 8.8/10 &  &  &   &  &  &  \\ 
10 & BAT & $9.3\pm1.0$ & 20.7/21 &  &  & & $0.3\pm1.6$ & $37.8\pm4.7$ & 19.6/20 \\ 
11 & BAT & $7.9\pm0.9$ & 9.3/11 &  &  &   &  &  &  \\ 
12 & BAT & $11.3\pm1.0$ & 18.2/22 &  &  &   & $-0.5\pm0.8$ & $44.1\pm5.7$ & 12.6/21 \\ 
13 & BAT & $13.6\pm2.5$ & 7.0/11 &  &  &   &  &  &  \\ 
14 & BAT & $11.2\pm2.4$ & 48.4/56 &  &  &   &  &  &  \\ 
15 & GBM & 10.2$\pm$0.5 & 79/67 & 6.4$\pm$1.9 & 14.2$\pm$3.2 & 56/64  & -0.2$\pm$0.5 & 40.0$\pm$3.6 & 47/66 \\ 
16 & BAT & $6.6\pm1.4$ & 9.6/7 &  &  &   & &  &  \\ 
17 & BAT & $10.1\pm1.0$ & 12.8/4 &  &  &   &  &  &  \\ 
18 & BAT & $16.8\pm3.2$ & 8.7/9&  &  &   &  &  &  \\ 
19 & BAT & $12.6\pm1.9$ & 11.2/10 &  &  &   &  &  &  \\ 
20 & BAT & $12.2\pm0.8$ & 35.8/25 &  &  &  & $-0.8\pm0.6$ & $50.5\pm7.6$  & 25.7/24  \\
20 & GBM & 12.4$\pm$0.7 & 83/66 & 7.9$\pm$1.9 & 19.0$\pm$4.6 & 83/64 & -0.3$\pm$0.4 & 53.0$\pm$5.2 & 67/65 \\ 
21 & BAT & $16.0\pm1.0$ & 50.5/33 & $3.4\pm0.8$ & $18.5\pm1.5$ & 27.3/31 & $-0.8\pm0.5$ & $78.3\pm25.6$ & 35.6/32 \\
21 & GBM & 15.6$\pm$0.7 & 80/66 & 2.8$\pm$0.8 & 17.6$\pm$1.2 & 55/64 & -0.1$\pm$0.3 & 68.5$\pm$6.8 & 57/65 \\ 
22 & BAT & $11.0\pm1.2$ & 10.2/8 &  &  &   &  &  &  \\ 
23 & BAT & $13.7\pm1.5$ & 6.0/5 &   &  &   & $-0.3\pm0.9$ & $60.1\pm16.9$ & 2.3/4  \\ 
24 & BAT & $15.7\pm0.8$ & 30.5/24 &   &  &   & $0.3\pm0.5$ & $66.0\pm5.7$ & 24.6/23  \\ 
24 & GBM$^{\diamond}$ & 19.6$\pm$1.6 & 72/66 & 5.1 & 21.5$\pm$2.3 & 51/65 & 0.4$\pm$0.7 & 82.0$\pm$12.0 & 66/65 \\ 
25 & BAT & & & $4.5\pm1.0$ & $35.8\pm18.0$ & 0.7/5 & & & \\ 
26 & GBM & 14.7$\pm$0.6 & 71/66 & 3.7$\pm$1.0 & 16.7$\pm$1.0 & 51/64 & -0.2$\pm$0.3 & 60.6$\pm$4.6 & 47/64 \\ 
27 & GBM & 7.3$\pm$1.1 & 79/66 & 4.6$\pm$1.3 & 23.1$\pm$8.2 & 77/64 & -1.9$\pm$0.8 & 29.7$\pm$101 & 54/65 \\ 
\hline
1  & BAT  & - & - & - & - & - & -1.0 & 33.2$\pm$5.5 & 16.7/21 \\ 
10 & BAT  & - & - & - & - & - & -1.0 & 34.6$\pm$6.5 & 21.6/21 \\ 
12 & BAT  & - & - & - & - & - & -1.0 & 42.1$\pm$7.0 & 13.4/22 \\ 
15 & GBM  & - & - & - & - & - & -1.0 & 39.0$\pm$5.3 & 47/67 \\ 
20 & BAT  & - & - & - & - & - & -1.0 & 50.8$\pm$7.0 & 25.9/25 \\ 
20 & GBM  & - & - & - & - & - & -1.0 & 56.7$\pm$8.7 & 56/66 \\ 
21 & BAT  & - & - & - & - & - & -1.0 & 84.3$\pm$15.8 & 35.8/33 \\ 
21 & GBM  & - & - & - & - & - & -1.0 & 86.6$\pm$14.5 & 63/66 \\ 
24 & BAT  & - & - & - & - & - & -1.0 & 43.8$\pm$10.6 & 3.7/8 \\ 
26 & GBM  & - & - & - & - & - & -1.0 & 71.0$\pm$9.5 & 52/66 \\ 
27 & GBM  & - & - & - & - & - & -1.0 & 49.0$\pm$15.4 & 55/66 \\ 

\enddata
\end{deluxetable}

We find that the bursts detected during the 2011 activity episode (Burst IDs 1 through 15) are best represented by a single BB function, with temperatures ranging between 4.5 and 15~keV. The fluences for these bursts were found to be quite low, mostly below 10$^{-8}$~erg~cm$^{-2}$. Three of these 15 events were also fit with the COMPT model, yielding slight improvements in the $\chi^2$ compared to the single BB model. However, the improvement in $\chi^2$ is insignificant given the introduction of an additional model parameter. Moreover, the PL index $\alpha$ from the COMPT fits for these three bursts could not be constrained (see Table~\ref{tab:spec}). The only recorded burst from 2012 (Burst ID 16), was also found to be rather dim, with a fluence of 5$\times$10$^{-9}$ erg cm$^{-2}$. The burst spectrum is well fit by a single BB model, with a temperature of 6.6~keV.

The 2015 re-activation of \xsrc commenced with three weak events (Burst IDs 17-19), and proceeded with much more energetic bursts. The spectra of the three weak events could also be modeled with a BB function. However, the spectral properties of the brighter events (Burst IDs 20-21 and 24-27) were best described with a BB+BB model with temperatures of $3-4$\,keV and $17-20$\,keV. The COMPT model fits to some of these events were statistically acceptable, but yielded poorly constrained or unconstrained model parameters ({\it i.e.}, the photon index, $\alpha$). Further analysis fixed the photon index to be -1, effectively turning the COMPT function into the functional form of the optically-thin thermal bremsstrahlung model. This resulted in the spectral cut-off energy parameter (E$_p$) being better constrained, with values varying between 30 and 50~keV (see Table~\ref{tab:spec}). 

In order to better constrain the parameters from the spectral fitting of the data using the aforementioned models, a joint-stacked analysis was applied to all ten \emph{Fermi}/GBM events. In addition to better constraining the spectral parameters through the minimization of the background using the limited amount of data available (one detector per outburst episode satisfies the $\theta$$\leq$40$^{\circ}$ criterion). This analysis also served to independently verify the \emph{Swift}/BAT spectral results of the same events, which were analyzed using XSPEC. Using an energy range of 8-200~keV and 8~ms time resolution, the BB+BB model was found to fit the combined data best. The BB+BB model fit to the stacked data, resulted in BB temperatures of $kT_{1}$ = 3.9$\pm$0.6 keV and $kT_{2}$ = 16.6$\pm$0.7 keV. The COMPT model fit the spectrum nearly well, yielding the PL index of $\alpha$=-0.29$\pm$0.18 and E$_{p}$=60.9$^{+2.9}_{-2.6}$ keV.

The joint spectral analysis of the BAT bursts observed in 2015, also found the combined BB+BB model to best fit the data ($\chi^2$/dof = 56.2/54). The BB temperatures were determined to be $kT_{1}$ = 3.9$\pm$0.7 keV and $kT_{2}$ = 16.9$\pm$0.8 keV, which are in perfect agreement with the results of the GBM joint spectral analysis. For the joint spectra of the BAT events, the COMPT and BB models performed worse with a $\chi^2$/dof = 64.6/55 and $\chi^2$/dof = 105.2/56, respectively. The joint spectrum for all of the weak bursts detected by BAT in 2011, is equally well described with the COMPT ($\chi^2$/dof = 46.1/55) and BB+BB models ($\chi^2$/dof = 44.8/54). For the former model, $\alpha$ was found to be unconstrained (-0.31$^{+0.33}_{-0.36}$), while the temperatures of the latter model were well constrained: $kT_{1}$ = 7.3$\pm$1.2~keV and $kT_{2}$ = 17.1$^{+4.1}_{-2.3}$~keV.

\section{Discussion}
\label{sect:discuss}

We have compiled the most comprehensive burst sample of \xsrcnos, comprising 27 bursts from its three burst active episodes in 2011, 2012 and the latest one in 2015 observed with \emph{Swift}/BAT and \emph{Fermi}/GBM: We have enhanced the number of bursts from \xsrc by about six-fold compared to what was previously observed  (Gavrill et al. 2011). We discuss below characteristic properties of these bursts in relation to the bursts from other magnetars, the persistent emission behavior of the source, as well as the properties of extended tail emission we identified with \emph{Swift}/XRT.

\subsection{Burst Properties}

The morphological properties of bursts from \xsrc are similar to typical magnetar events. They all have a duration ranging from 4 to 700~ms, with more than 80~\% of the bursts detected by BAT lasting $\leq$ 50~ms. Although this sample size is too small to make any definitive conclusions regarding the duration of all bursts from this source, they do appear to be shorter than the burst durations of other magnetars (see e.g., G{\"o}{\v g}{\"u}{\c s} et al. 2001, Gavriil et al. 2004, van der Horst et al. 2012).

The spectra of bursts from \xsrc exhibit diverse characteristics. Relatively dim bursts, with fluences less than 8$\times$10$^{-9}$~erg~cm$^{-2}$, and almost all events observed during the 2011 activity episode, were best represented with a single black-body (BB) function. This is quite similar to what has been observed for the October 2008 bursts of SGR J1550-5418 (von Kienlin et al. 2012) as well as the bursts from the first transient magnetar, XTE J1810-197. The bursts of the 2015 active episode are comparatively brighter and their spectral shapes are statistically better represented with more complex models, such as the sum of two black-bodies (BB+BB) or COMPT. The dilemma of whether typical magnetar burst spectra are predominately thermal (BB+BB) or non-thermal (COMPT), is largely unresolved (Lin et al. 2011, van der Horst et al. 2012). The thermal scenario of magnetar bursts may be indicative of emission originating from the neutron star surface due to either energy dissipated in the crust or surface heating by return currents from twisted magnetic field lines (see {\it e.g.}, Beloborodov \& Thompson 2007), while the non-thermal model implies magnetospheric processes are more dominant. In reality, what unfolds in the vicinity of highly-magnetized systems is likely to be more complicated, and both surface and magnetospheric processes could be coupled together as a consequence of these environmental conditions. 

We find that the peak energy parameters of the COMPT model were of the order of or larger than 50~keV. This is in agreement with what has been measured for dim bursts from SGR 0501+4516 (Lin et al. 2011) and SGR J1550$-$5418 (van der Horst et al. 2012). When the spectra are fit with a special case of the COMPT model, namely when its PL index is fixed to -1, we obtained statistically acceptable fits to nearly all brighter bursts, and the peak energy was of the order of 40$-$50~keV, or less. The temperatures of the BB+BB fits were around 3$-$4~keV and 17~keV, similar to the spectral characteristics of bursts from other magnetars. Broad-band spectral coverage is required to conclusively determine which of these models best represents the magnetar spectra (Lin et al. 2013).

\subsection{Outburst Properties}

The three burst-active episodes from \xsrc presented here are not the only ones from this source. Short bursts from \xsrc were detected in monitoring observations of the source with RXTE in 2006 and 2007 (Gavriil et al. 2011). A total of six bursts were reported, four of which were seen within about 4.5 minutes on 2006 June 25. None of these six events were able to trigger the BAT, which was the only wide-area sky-monitoring satellite at that time. 

Magnetar \xsrc burst-active episodes resemble that of magnetars with low-bursting rates (with one or a few bursts per reactivation), such as SGR 0418+5729 (van der Horst et al. 2010), and SGR 1833$-$0833 (G{\"o}{\v g}{\"u}{\c s} et al. 2010). However, there is an important difference between \xsrc and the latter, namely, their outbursts usually lead to long-lasting (months to years) flux enhancements (Rea \& Esposito 2011), while those of \xsrc are not observed to cause any significant long-lasting flux enhancements, similar to the behavior observed with 1E 1841$-$045 (Lin et al. 2013).

\xsrc undergoes an activity episode on a time-scale that ranges from several months to a few years. According to the magnetar model, the neutron star crust is stressed by the diffusion of the strong magnetic field which drives it to a critical strain, at which a slight disturbance could fracture the crust and give rise to energetic bursts (Thompson \& Duncan 1995, Thompson, Yang and Ortiz 2016). This so-called self organized criticality (SOC), has been shown to occur in magnetar bursts (G{\"o}{\v g}{\"u}{\c s} et al. 1999, 2000; Gavriil et al. 2004, Scholz \& Kaspi 2011). The cluster of bursts seen in the 2011 and 2015 reactivation of \xsrc also suggests the SOC scenario: the impact of a leading burst in an activity phase brings the strain of nearby crustal sites to the level of criticality, quicker than their natural progression under internal and external magnetic stresses. Note the important fact that the SOC behavior is also expected to occur in the magnetic reconnection process (Aschwanden et al. 2016).

\subsection{Extended Burst Tail Emission}

The initial burst that triggered \emph{Swift}/BAT on 2015 February 28, led to the detection of the decaying flux enhancement or extended tail of the source emission. Similar burst tails have been observed from other magnetars, such as; SGR 1900+14 (Lenters et al. 2003), SGR 1806-20 (G{\"o}{\v g}{\"u}{\c s} et al. 2011), SGR 1550-5418 ({\c S}a{\c s}maz Mu{\c s} et al. 2015), as well as from \xsrc in 2006 (Gavriil et al. 2011, Chakraborty et al. 2016). More recently, such a tail was identified following a burst from a rotation powered pulsar, PSR J1119$-$6127 (G{\"o}{\v g}{\"u}{\c s} et al. 2016). The leading bursts in such events tend to be more energetic, however, an energetic event does not necessarily mean that an extended tail will be present. The X-ray spectra of all the extended tails exhibited thermal signatures. The tail discovered in this study is no exception; the BB temperature declined from 1.6~keV to about 1~keV over the course of about 300~s. This is in line with previously observed extended burst tails (see {\it e.g.}, Lenters et al. 2003, G{\"o}{\v g}{\"u}{\c s} et al. 2011), indicating thermal cooling of burst-induced phenomena.

X-ray pulsation properties of the underlying neutron stars are usually affected by these events. In particular, the pulsed amplitude of their X-ray emission was enhanced in the burst tails (see related references cited above). We present clearly noticeable pulsations from \xsrc during the tail in Figure~\ref{bat_xrt_lc}. Amplified pulsations were also the case in its 2006 extended burst tails (Gavriil et al. 2011). It is possible that the leading burst in this event caused a trapped fireball, similar to the agent responsible for the oscillating tails of giant flares (Thompson \& Duncan 1995; 2001), but on a much smaller scale. The periodic modulations are then naturally observed when the cooling fireball came into the field of view of \emph{Swift}. It is important to note that sustaining an optically thick pair plasma fireball would require much higher temperature than about 1 keV. However, the required energy budget might not necessarily need to be supplied from the burst, but might be readily available, as \xsrc is a persistent emitter of bright hard X-rays. Alternatively, a burst-induced heating of a portion of the neutron star surface, a hotspot, could also account for the enhanced pulsations in the decaying X-ray flux enhancement.

We found that many bursts during the flux enhancement align near the maximum of the neutron star spin phase. Extensive investigations for phase alignment of magnetar bursts were mostly inconclusive, except for XTE J1810-197: in that case energetic bursts (identified as spikes of $0.5-2.0$\,s duration) were seen separated from each other by 5.54 s, the spin period of the source (Woods et al. 2005). However, peak-phase aligned X-ray bursts are difficult to accommodate, either with a localized fireball scenario or with a hotspot. It is possible that they might arise from a different mechanism at a similar location on the neutron star surface, such as the leading burst driven  instabilities causing small scale magnetar bursts near the magnetic pole of the neutron star. Next generation space telescopes with large collecting area and X-ray polarimetry capability could solve the puzzle of whether the enhanced X-ray pulsations and peak aligned X-ray bursts are driven by the same mechanism, or they are caused by somehow associated and spatially coincident different physical phenomena.

\section{Conclusions}

\xsrc is an active magnetar. Besides its persistent emission of radiation from infrared to hard X-rays, it is also emitting energetic bursts. Its unpredictable burst active episodes repeat on a timescale from about six months to $\sim4.5$ years. Bursts from this source morphologically resemble typical short bursts from magnetars, but less energetic compared to the bulk of magnetar bursts. The extended burst tail emission following a burst on the 2015 February 28, has a thermal nature, cooling over a time-frame of several minutes. This behavior is similar to what has been observed previously from other magnetars, as well as from \xsrc itself in 2006. Finally, we uncovered phase aligned X-ray bursts/spikes during the 2015 extended burst tail, which are likely associated to a contemporaneous but different physical phenomenon. 

\acknowledgments

We would like to thank the anonymous referee for very constructive comments. E.G. and Y.K. acknowledge support from the Scientific and Technological Research Council of Turkey (T\"UB\.ITAK, grant no: 115F463). L.L. is supported by the Fundamental Research Funds for the Central Universities and the National Natural Science Foundation of China (grant no. 11543004). O.J.R. acknowledges support from Science Foundation Ireland under Grant No.~12/IP/1288. C.K. and G.Y. acknowledge support from NASA grant NNH07ZDA001-GLAST (PI: C. Kouveliotou).

\end{document}